# Predicting Nonfarm Employment

Tarun Bhatia
Applied Artificial Intelligence
Research & Georgia Institute
Of Technology
USA
tbhatia30@gatech.edu

## ABSTRACT

U.S. Nonfarm employment is considered one of the key indicators for assessing the state of the labor market. Considerable deviations from the expectations can cause market moving impacts.

In this paper, the total U.S. nonfarm payroll employment is predicted before the release of the BLS employment report. The content herein outlines the process for extracting predictive features from the aggregated payroll data and training machine learning models to make accurate predictions.

Publically available revised employment report by BLS is used as a benchmark. Trained models show excellent behaviour with $R^2$ of 0.9985 and 99.99% directional accuracy on out of sample periods from January 2012 to March 2020.

## 1. INTRODUCTION

Even for the best investors, delivering consistent returns in excess of the market is challenging. The Bureau of Labor Statistics (BLS) collects data each month on employment, hours, and earnings from a sample of nonfarm establishments through the Current Employment Statistics (CES) program. The CES survey includes about 145,000 businesses and government agencies, which cover approximately 697,000 individual worksites drawn from a sampling frame of Unemployment Insurance (UI) tax accounts covering roughly 10.2 million establishments. The active CES sample includes approximately one-third of all nonfarm payroll employees in the 50 states and the District of Columbia.

BLS publishes "Total NonFarm Payroll", an important publication from the survey that moves the market. This is one of the first indicators of the labor market and how well the economy is doing. Significant changes than expected results in huge market movements, hence it is an important indicator for traders and investors.

The BLS releases a monthly Employment Situation Report every month. It is usually released on the third Friday after the conclusion of the reference week, which is the week that includes the 12th of the month.

Payroll transactional data along with natural disaster and weather information is used to forecast total U.S. Nonfarm payroll employment numbers given by BLS before their release date. Our out of sample historical forecasts are very close to that of BLS with a R squared of 0.9985 and directional accuracy of 0.9899 over the period of Jan 2012 to March 2020. Armed with this information traders and investment can make much better market predictions before the release of CES employment situation report..

## 2. RELATED WORK

There have been many works in estimating Nonfarm employment. BLS estimates this by giving out two surveys, which are monthly. One is the CES survey and other is the Current Population Survey (CPS) survey which surveys a sample of about 60,000 households. CES surveys are usually considered more accurate. ADP estimates monthly nonfarm payroll employment in their monthly National Employment Report (NER) report which is released two days earlier than the release of BLS report. Various analysts and trading professionals try to predict this, but we could not find any published paper describing the methodology for those predictions.

## 3. DATA OVERVIEW

### 3.1 Private Payroll Datasets

Private payroll transactional data is collected from top 5 leading payroll providers in the U.S. The Dataset is aggregated monthly based on company size, industry sectors, tenure, job type. Historically the payroll providers may have grown and acquired new clients. To prevent bias from its growth, a two month rolling period is used, where the number of companies/employers remains constant.

### 3.2 Natural Disaster Dataset

Storm Data[2] is provided by the National Weather Service (NWS) and contain statistics on personal injuries and damage estimates to business. The data contain a chronological listing, by state, of hurricanes, tornadoes, thunderstorms, hail, floods, drought conditions, lightning, high winds, snow, temperature extremes and other weather phenomena. It is publicly available at https://data.nodc.noaa.gov

### 3.3 Weekly Unemployment Insurance claims Dataset

Every week, the Department of labor releases its Unemployment weekly claims report (Sunday to Saturday). The weekly data is aggregated to monthly. Initial claims measure emerging unemployment and continued weeks claimed measure the number of persons claiming unemployment benefits.

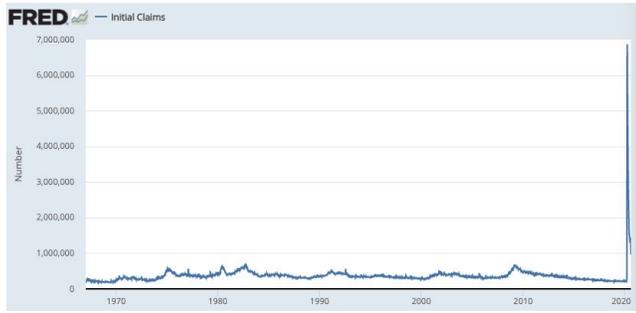

## 4. METHODOLOGY

Analysis of the payroll datasets is accomplished through a series of sequential steps. The first step focuses on evaluating data integrity and visualizing raw data. This analysis is then used to determine the raw features used to create meaningful derived features. Further, derived features in conjunction with machine learning algorithms predict actual employment with BLS revised report as the benchmark.

### 4.1 Validate Datasets

The first task is to verify the cleanliness and integrity of the aggregated payroll data, and resolve any potential issues. This includes checking data for outlier values, data type consistency, as well as consistency against the data dictionary. Additionally, the dataset is checked for presence of unexpected missing values..

### 4.2 Payroll vs CES report

We will discuss some of the differences between the BLS and the Payroll data.

a. Even though we aggregated data through leading payroll providers, the composition of the companies paid by payroll providers does not match the size and industrial composition of the companies in the general population. This effect is even more pronounced across the various sectors.

b. Payroll providers' employment number includes all active employees in a month whereas BLS employment survey only tracks the employees who were actually paid during the reference month which includes the 12 th of the month.

c. Payroll providers may acquire or lose clients as a company historically hence, there is a bias in their data due to their growth as a company compared to the general population.

d. Payroll providers' coverage for small businesses and some of the cyclical sectors is not representative of the general population.

e. BLS surveys are not accurate and contain errors due to delay in responses. In some months like August, managers usually go on leave and hence do not respond to the surveys timely. Errors may also come from the businesses which were shut down in between the survey period.

f. Both payroll providers and employers may double count the employment, if people are doing multiple jobs.

g. BLS CES survey and payroll data, both have their unique sampling biases. BLS tries to account for these sampling biases using their methods described in the CES report, but those are still based on assumptions and not exactly accurate.

Since our goal is to predict the BLS CES survey total nonfarm payroll employment, we need to account for biases in the data due to reasons discussed above.

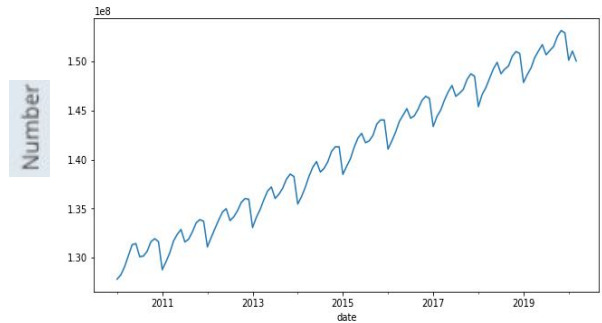

*Fig 1: Total nonfarm payroll employment vs BLS report*

### 4.3 Detrending

First, we see a deterministic linear upward trend in the BLS total nonfarm payroll employment. This Trend interferes with models to make accurate predictions. The effect is especially more pronounced for tree based models which we use later.

Therefore, we detrend the data by using the first order differencing to make it easier for the model to predict and it gave the best results as compared to other methods for detrending.

$$value(t) = observation(t) - observation(t-1)$$

Intuitively, the series has now reduced to the change in the total BLS nonfarm payroll employment between the consecutive months.

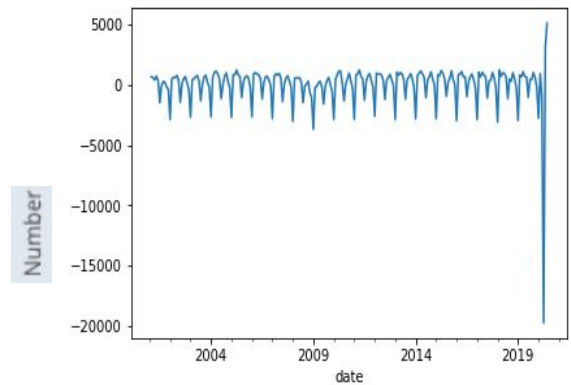

*Fig 2: Detrended Total nonfarm payroll employment vs BLS report*

## 4.4 Deseasoning

Even after detrending, we see a clear seasonal pattern in the time series. We try two different approaches to modelling and compare:

a. Removing seasonality from both our input features and BLS target variables
b. Keeping the seasonality in input features and BLS target variables.

There are various pros and cons to each approach, we will discuss these here.

Seasonality often interferes with model predictions just like trends. This effect is more pronounced especially when series for input features have different seasonality than output targets. Therefore, for the first approach we remove seasonality by using first order differencing with 12 periods (12 months, assuming yearly seasonality).

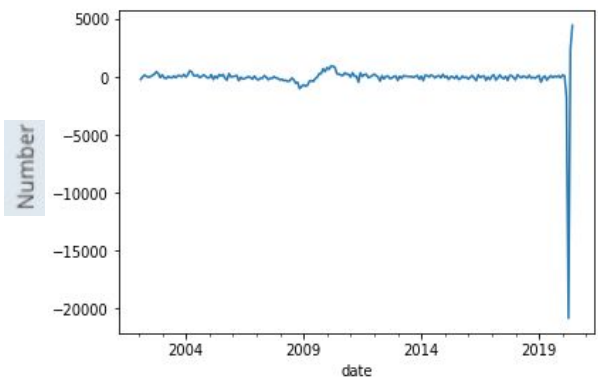

*Fig 3: Deseasoning Total no farm payroll employment vs BLS report*

## 4.5 One hot encoding

As described earlier, in order for a model to learn the seasonal information, monthly information is one hot encoded. Normal integer encoding tends to cause poor performance and unexpected results as the model assumes the natural order, in our case for months. Let's assume we encode January as 1 and December as 12. Even though these two are consecutive months, their natural ordering makes them far apart compared to others. This may cause unwanted behaviour.

We express One-hot encoding formally here. Let x be the discrete categorical variable with n distinct values $x_1, x_2, ... x_n$. Then, the One-hot encoding of a particular value $x_i$ is a vector v where the ith component has the value 1 and all other components have the value zero. In our case, a variable representing months takes values from the set S = 1 to 12 . Let $x_1 = 1, x_2 = 2,$ and $x_3 = 3$.

A One-hot encoding for x is: **(1, 0, 0, 0, 0, 0, 0, 0, 0, 0, 0, 0)**, **(0, 1, 0, 0, 0, 0, 0, 0, 0, 0, 0, 0)**, and **(0, 0, 1, 0, 0, 0, 0, 0, 0, 0, 0, 0)**.

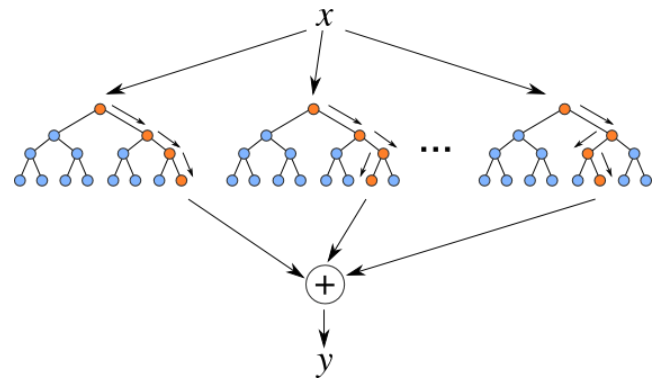

*Fig 4: One hot encoding example*

## 4.6 Modeling

We preprocessed BLS total nonfarm payroll employment numbers as targets in a supervised learning setting and add back the trend later. An ensemble of Extremely Randomized Trees proposed by Geurts et al. (2006) is used for this regression problem. The reason for choosing the ensemble of extremely randomized trees is primarily due to its ability to use multitude of features in a non linear model without overfitting. This is very important to us as we are working with a relatively smaller number of datapoints. Additionally, they are also very competitive in computing times and offer an insight into the model using feature importances.

*Fig 5: Extremely randomized Tree depiction*

We refer the reader to (Geurts et al., 2006) for a more formal description of the algorithm and a detailed discussion of its main features, but we will describe extremely randomized trees here in a nutshell.

An ensemble of Extremely Randomized Trees builds multiple trees and splits nodes using random subsets of features, similar to random forests but has some key differences. Firstly, it does not bootstrap observations, or in other words it samples without replacement. Secondly, for each of the features (randomly selected at each interior node) we randomly choose a discretization threshold to define a split, instead of choosing the best split based on the local sample.

## 4.7 Backtesting using expanding window

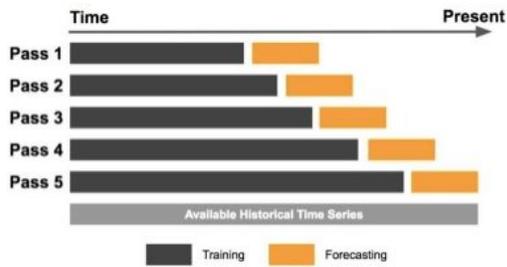

*Fig 6: Expanding window backtest*

A model may provide good in-sample fit to historical data and produce forecasts that differ significantly from actual values (ex ante). To test a model's performance on time series data we often require chronological testing. This chronological testing is referred to as backtesting. A common backtesting method is to split the time series data into fixed training and testing split while preserving chronology. However this method prevents the model from using recent information to train.

Since we have fewer data points, we choose to use an expanding window based backtesting approach[3]. In this approach, the data is split into a number of training and test splits chronologically. Starting at the beginning of the time series, a minimum number of samples in the window are used to train a model. The model makes a prediction for a fixed window of timesteps. The forecasts are evaluated against the actual values and stored. Then, the window is expanded to include the forecasted values, and the process is repeated.

## 5. RESULT AND ANALYSIS

For both the approaches, we found data to be highly predictive and achieved high R squared and directional accuracy out of sample on the expanding windows. Extremely randomized tree ensembles helped us prevent overfitting, as being one of our major concerns with smaller number of datapoints and variety of features.

For the first approach using both detrending and deseasoning on both Payroll and BLS CES survey data, we found that the model performed exceptionally well, with $R^2$ of 0.998509 and directional accuracy of 0.9694 when compared to benchmark which was BLS CES reported data for Total payroll nonfarm employment.

For the other approach, where we detrend the targets and features and use seasonal features as one-hot encoding, the model achieved higher directional accuracy of 0.9899 with slightly reduced R2 of 0.99846.

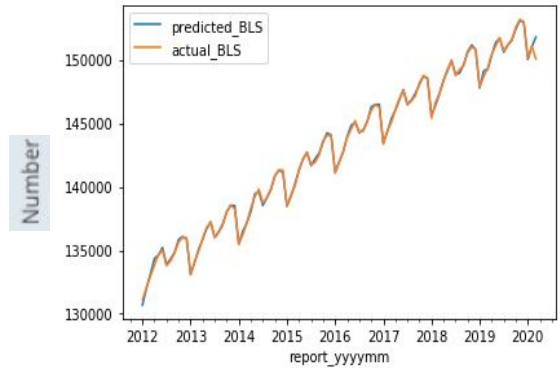

*Fig 7: Predicted total Nonfarm payroll employment vs actual reported nonfarm payroll using approach 1*

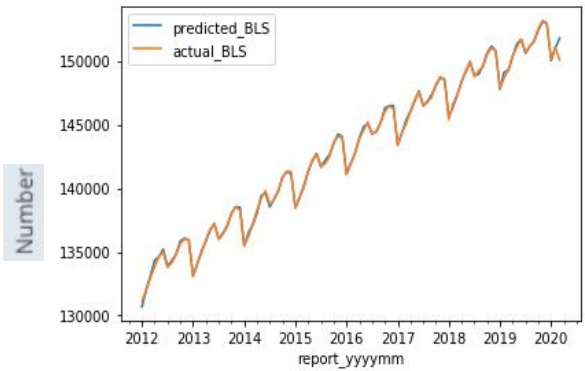

*Fig 8: Predicted total Nonfarm payroll employment vs actual reported nonfarm payroll using approach 2*

In both the approaches, the predictions for March 2020 were far from accurate. This has severely impacted the overall high R squared and directional accuracies on out of sample for both the models. As noted earlier there is a difference between the way payroll providers track employment vs BLS tracks employment. Analysing payroll data further, we found that the drop in employment is not significantly lower for predictions to be closer to BLS employment numbers. This may be due to the difference in the data collection methodology.

Payroll providers employment numbers include all active employees in a month from start of the month until the end of the month, whereas BLS employment survey only tracks the employees who were actually paid during the reference month which includes the 12th of the month. During normal circumstances, our models were able to adjust that with weights, but during extraordinary circumstances like pandemic, these weights were not reflective of actual change, In order to account for that essentially, we would like to have more granularity in the data.

# 6. CONCLUSION

Payroll data was validated and analysed for predicting total non-farm payroll employment. Our trained models were able to achieve excellent R squared scores and near to perfect directional accuracy in out of sample with respect to the BLS CES report as the benchmark. This is quite encouraging for the fact that payroll providers' data collection methodology differs from BLS's data collection methodology.

In the Payroll data, the employment numbers include all active employees in a month, whereas BLS employment survey only tracks the employees who were actually paid during the reference month which includes the 12 th of the month, a number of factors like labor strike, weather impact, natural disasters and pandemic can lead to significant difference in these numbers. People may be actively listed as employees but may not get paid during the reference month that includes 12th. In this research we tried to incorporate natural disaster, storm and weather dataset to account for that. There is still scope for improvement by incorporating other orthogonal datasets.

We can take our research one step further by looking at the effect of surprise between our model's predictions and analysts reports, and use that as a macro indicator to predict stock market's performance.

Based on these predictions, we can also create sector rotation based model portfolios.